\documentclass[sigconf]{acmart}

\usepackage{multirow}
\usepackage{graphicx}
\usepackage{makecell}
\usepackage{subcaption}
\usepackage{caption}
\usepackage{enumitem}
\usepackage{array}
\usepackage{longtable}
\usepackage{xcolor}
\usepackage{booktabs}
\usepackage{colortbl}
\definecolor{mygray}{gray}{0.9}
\usepackage{ltablex}
\usepackage{supertabular}

\keepXColumns

\AtBeginDocument{%
  }

\copyrightyear{2026}
\acmYear{2026}
\setcopyright{cc}
\setcctype{by}
\acmConference[KDD '26]{Proceedings of the 32nd ACM SIGKDD Conference on Knowledge Discovery and Data Mining V.2}{August 09--13, 2026}{Jeju Island, Republic of Korea}
\acmBooktitle{Proceedings of the 32nd ACM SIGKDD Conference on Knowledge Discovery and Data Mining V.2 (KDD '26), August 09--13, 2026, Jeju Island, Republic of Korea}
\acmDOI{10.1145/3770855.3818911}
\acmISBN{979-8-4007-2259-2/2026/08}

\begin{document}

\title[AnyPPG: An ECG-Guided PPG Foundation Model for Holistic Health Profiling]{AnyPPG: An ECG-Guided PPG Foundation Model Trained on Over 100,000 Hours of Recordings for Holistic Health Profiling}

\author{Guangkun Nie}
\affiliation{%
  \institution{School of Intelligence Science and Technology, Peking University}
  \city{Beijing}
  \country{China}
}
\additionalaffiliation{%
  \institution{State Key Laboratory of General Artificial Intelligence, Peking University}
  \city{Beijing}
  \country{China}
}
\email{nieguangkun@stu.pku.edu.cn}

\author{Xiaocheng Fang}
\authornotemark[1]
\affiliation{%
  \institution{School of Intelligence Science and Technology, Peking University}
  \city{Beijing}
  \country{China}
}
\email{fangxiaocheng162@gmail.com}

\author{Gongzheng Tang}
\affiliation{%
  \institution{National Institute of Health Data Science, Peking University}
  \city{Beijing}
  \country{China}
}
\email{gztang@hsc.pku.edu.cn}

\author{Yujie Xiao}
\affiliation{%
  \institution{National Institute of Health Data Science, Peking University}
  \city{Beijing}
  \country{China}
}
\email{xiaoyujie@stu.pku.edu.cn}

\author{Jun Li}
\affiliation{%
  \institution{National Institute of Health Data Science, Peking University}
  \city{Beijing}
  \country{China}
}
\email{nickljlee0306@gmail.com}

\author{Bo Liu}
\authornotemark[1]
\affiliation{%
  \institution{School of Intelligence Science and Technology, Peking University}
  \city{Beijing}
  \country{China}
}
\email{liubo2022@stu.pku.edu.cn}

\author{Hongyan Li}
\authornotemark[1]
\authornote{Corresponding authors.}
\affiliation{%
  \institution{School of Intelligence Science and Technology, Peking University}
  \city{Beijing}
  \country{China}
}
\email{leehy@pku.edu.cn}

\author{Shenda Hong}
\authornotemark[2]
\affiliation{%
  \institution{National Institute of Health Data Science, Peking University}
  \city{Beijing}
  \country{China}
}
\email{hongshenda@pku.edu.cn}

\renewcommand{\shortauthors}{Guangkun Nie et al.}

\begin{abstract}
Photoplethysmography (PPG) is widely used as a non-invasive and accessible modality for continuous health monitoring. However, despite being a peripheral hemodynamic signal intrinsically coupled with systemic circulation, existing research has largely confined its scope to a narrow range of cardiovascular tasks, leaving a fundamental question underexplored: to what extent can PPG support holistic health profiling beyond traditional cardiovascular applications? To answer this question, we present AnyPPG, a foundation model-based framework designed to reveal the broader health-profiling potential of PPG. To ensure reliable performance for this investigation, AnyPPG is pretrained with ECG guidance on the most diverse PPG corpus with synchronized ECG to date, comprising over 100,000 hours of recordings from six large-scale data sources. This pretraining yields robust and physiologically grounded PPG representations that provide a reliable basis for subsequent analysis. Building upon this pretrained model, we conduct a systematic investigation into the association between PPG and holistic health through, to the best of our knowledge, the first PPG-based phenome-wide disease detection study, spanning 1,468 disease phenotypes in more than 15,000 subjects. Our evaluation demonstrates the effectiveness of AnyPPG: across eight clinical and wearable datasets covering 15 downstream tasks, it achieves the best performance in 13 tasks. More importantly, in the phenome-wide analysis, AnyPPG exhibits meaningful discriminative capability (AUC $\ge$ 0.70) for 307 phenotypes across 16 distinct phecode chapters, including 230 non-circulatory conditions such as dementia, chronic kidney disease, hyperkalemia, and glaucoma, many of which have rarely been explored using PPG. Collectively, these findings indicate that easily acquired PPG signals encode rich health-related information extending well beyond conventional cardiovascular assessment, establishing a promising foundation for future research toward broader, scalable, and non-invasive PPG-based health applications. Code and model weights are available at \url{https://github.com/PKUDigitalHealth/AnyPPG}.
\end{abstract}



\maketitle

\section{Introduction}

\begin{figure*}
    \centering
    \includegraphics[width=\linewidth]{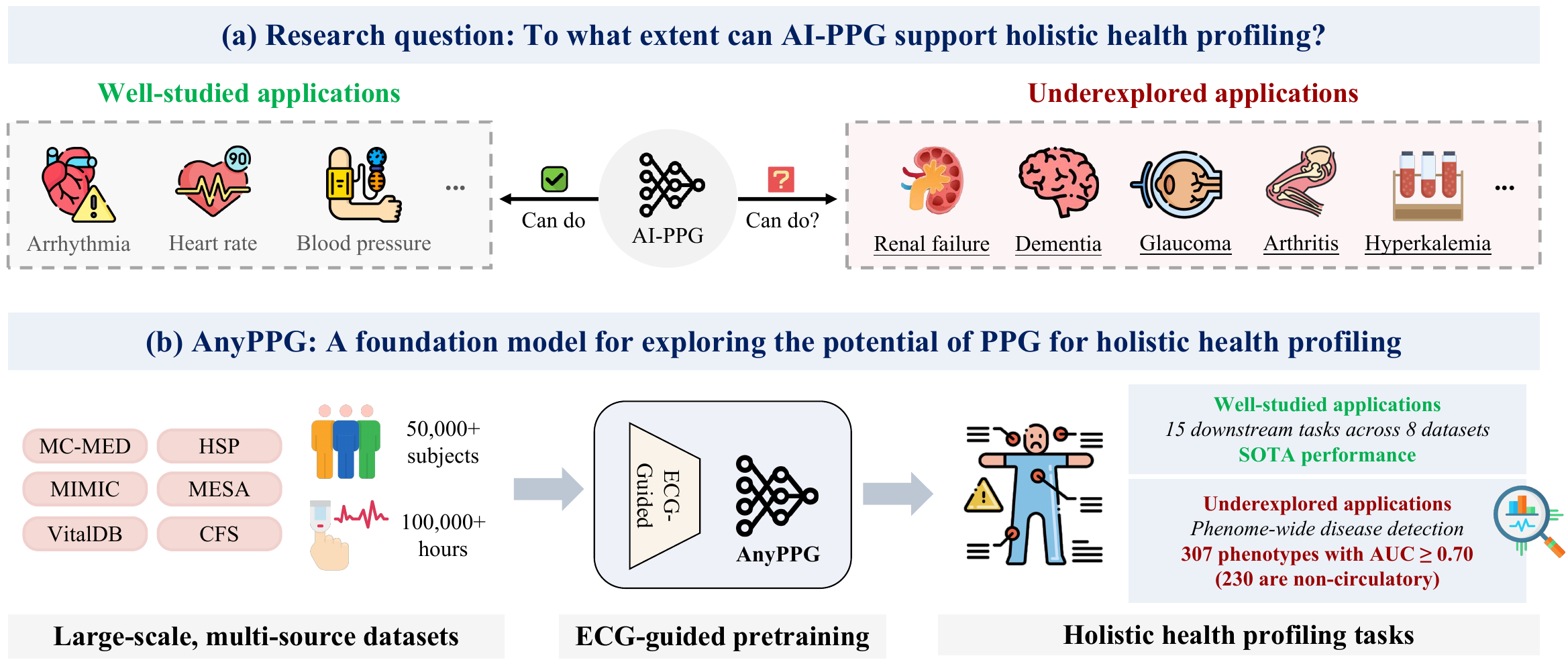}
    \caption{Research question and study overview of AnyPPG. (a) While PPG has been extensively studied for cardiovascular applications, its potential to support holistic health profiling across multi-organ conditions remains unclear. (b) To answer this question, we present AnyPPG, a foundation model-based framework designed to investigate the holistic health-profiling potential of PPG. AnyPPG leverages ECG-guided pretraining on the most diverse PPG corpus with synchronized ECG recordings to date, enabling the learning of physiologically grounded PPG representations and achieving SOTA performance on conventional PPG analysis tasks. Building upon this foundation, it further reveals broad associations between PPG and holistic health through a phenome-wide disease detection study spanning 1,468 phenotypes.}
    \label{fig:overview}
\end{figure*}

With the widespread adoption of wearable technologies, photoplethysmography (PPG) has emerged as a key modality for continuous, out-of-clinic health monitoring owing to its non-invasive nature and ease of integration \cite{bayoumy2021smart, williams2023wearable, spatz2024wearable}. 
PPG measures variations in peripheral blood volume, producing a waveform that reflects physiological attributes such as cardiac output, vascular compliance, 
and autonomic regulation \cite{nie2024review, jiang2025ppg}. Recent advances in artificial intelligence have further enabled the effective utilization of PPG (AI-PPG) across a broad spectrum of applications, such as heart rate estimation \cite{pankaj2022review, bent2020investigating}, cardiovascular disease screening \cite{elgendi2019use, pereira2020photoplethysmography}, and digital biomarker development \cite{nie2025artificial, miller2025wearable}.

Despite substantial progress, existing research on PPG has remained largely focused on a narrow set of cardiovascular-related tasks, leaving its broader physiological relevance underexplored. PPG originates from the circulatory system, which is tightly coupled with organs and tissues throughout the body, suggesting that PPG signals may encode physiological information beyond conventional cardiovascular assessment. In contrast, recent studies on electrocardiography (ECG), which reflects the electrical activity of the heart and is closely linked to PPG, have demonstrated that cardiac electrical signals can support health assessment across a far broader clinical scope, ranging from the identification of kidney and liver diseases \cite{holmstrom2023deep, lin2026ai, alcaraz2025electrocardiogram, simonetto2025detection} to full-spectrum analyses spanning diverse disease categories \cite{hughes2025deep, li2026anyecg}. This contrast raises a fundamental question: \textbf{to what extent can PPG support holistic health profiling beyond traditional cardiovascular applications?}

To bridge this research gap, we introduce AnyPPG, a foundation model-based framework designed to reveal the broader health-profiling potential of PPG. We organize this study under a two-stage paradigm. In the first stage, AnyPPG is pretrained on the most diverse PPG corpus with synchronized ECG to date, comprising over 100,000 hours of recordings from six large-scale data sources. This large-scale pretraining enables the model to learn robust and transferable PPG representations, providing a reliable foundation for subsequent investigation. Unlike prior approaches that rely solely on unimodal self-supervised learning \cite{abbaspourazad2023large, pillaipapagei, saha2025pulse, chen2025gpt, ding2024siamquality}, AnyPPG incorporates ECG as a source of cross-modal physiological supervision. Such supervision guides representation learning toward physiologically meaningful cardiovascular dynamics consistent with ECG while reducing sensitivity to non-physiological artifacts, yielding representations that are both robust and physiologically grounded. In the second stage, building upon this high-performing foundation model, we conduct a systematic investigation into the association between PPG and holistic health through a large-scale PPG-based phenome-wide disease detection study spanning 1,468 disease phenotypes in more than 15,000 subjects, encompassing conditions far beyond well-studied cardiovascular diseases.

We begin by benchmarking AnyPPG on conventional PPG analysis tasks to assess representation quality, where it consistently achieves state-of-the-art (SOTA) performance compared with unimodal self-supervised methods, general-purpose time-series foundation models, and specialized PPG foundation models across eight independent downstream datasets, attaining the best results in 13 of 15 tasks. Building upon this validated foundation, the phenome-wide analysis further reveals meaningful discriminative capability across 16 distinct disease phenotype categories, with 307 phenotypes achieving an area under the curve (AUC) of at least 0.70. Strong performance is observed for 77 circulatory disorders, such as congestive heart failure (AUC = 0.84) and rheumatic heart valve disease (AUC = 0.81). Importantly, comparable discriminative capability is also demonstrated for 230 non-circulatory phenotypes, including dementia (AUC = 0.81), chronic renal failure (AUC = 0.75), hyperkalemia (AUC = 0.75), and glaucoma (AUC = 0.72), many of which have rarely been explored using PPG in prior research.

Specifically, our contributions are threefold:
\begin{enumerate}[nosep]
\item We introduce AnyPPG, an ECG-guided PPG foundation model pretrained on the most diverse multi-source PPG corpus with synchronized ECG to date, yielding robust and transferable PPG representations.
\item We conduct a comprehensive evaluation of AnyPPG across eight diverse downstream datasets spanning both clinical and wearable settings, demonstrating consistent SOTA performance and achieving the best results in 13 of 15 tasks.
\item Building upon AnyPPG, we conduct, to the best of our knowledge, the first PPG-based phenome-wide disease detection study, spanning 1,468 disease phenotypes. This analysis provides quantitative evidence of broad associations between PPG and diverse disease phenotypes, many of which have rarely been explored in prior research, thereby revealing new directions for scalable and non-invasive PPG-based health assessment.
\end{enumerate}

\section{Related Work}
\subsection{PPG Foundation Models}
Foundation models have demonstrated strong potential in healthcare and have been widely studied across domains such as computational pathology \cite{xu2024whole, lu2024multimodal}, medical imaging \cite{ma2025fully, sun2025foundation, he2024exploring}, and physiological signal analysis \cite{li2025electrocardiogram, thapa2026multimodal, nie2025low}. Through large-scale pretraining with supervised or self-supervised objectives, these models aim to learn transferable representations that generalize across diverse downstream tasks. In the context of PPG, existing foundation models primarily exploit intrinsic signal characteristics, including inter-individual variability, waveform morphology, susceptibility to noise and artifacts, and temporal predictability. Abbaspourazad et al. \cite{abbaspourazad2023large} and PaPaGEI-P \cite{pillaipapagei} employ individual-level contrastive learning to encourage discrimination between PPG signals from different subjects, thereby capturing subject-specific representations. PaPaGEI-S \cite{pillaipapagei} and PulsePPG \cite{saha2025pulse} further incorporate waveform morphology by designing contrastive objectives based on physiological indices (e.g., stress-induced vascular response index) or motif-level similarity, enabling the extraction of physiologically meaningful features. Complementary to these approaches, SiamQuality \cite{ding2024siamquality} adopts a contrastive learning formulation that aligns representations of low- and high-quality PPG segments, improving robustness to signal degradation and noise. GPT-PPG \cite{chen2025gpt}, in contrast, explores generative pretraining inspired by autoregressive language models to learn representations from PPG time series. In parallel, a line of work on foundation models for wearable sensing leverages multiple sensor modalities, including PPG or its derived signals \cite{erturk2025beyond, narayanswamyscaling}. However, these approaches are not specifically designed for PPG, and therefore differ fundamentally in modeling objectives from PPG-focused foundation models.

To summarize, existing PPG foundation models primarily rely on unimodal self-supervised learning from PPG signals. In addition, their pretraining data are often derived from single-source \cite{abbaspourazad2023large, ding2024siamquality, chen2025gpt, saha2025pulse} or weakly heterogeneous datasets \cite{pillaipapagei}. 
While effective at capturing specific signal characteristics, this paradigm may constrain the robustness and transferability of the learned representations. 
Given growing evidence that multimodal representation learning can improve representation quality and generalization \cite{radford2021learning, yu2022coca, zhang2025sensorlm, guo2023siamaf}, incorporating multi-source data and cross-modal physiological supervision represents a promising direction for advancing PPG foundation models.

\subsection{PPG-Based Healthcare Applications}
As a physiological signal reflecting circulatory dynamics, PPG enables convenient and continuous monitoring of cardiovascular function and has been extensively studied in healthcare research. Existing PPG-based studies have predominantly focused on cardiovascular-related applications, encompassing the estimation of physiological parameters such as heart rate \cite{pankaj2022review, bent2020investigating, panwar2020pp}, systolic/diastolic blood pressure \cite{schlesinger2020blood, panwar2020pp}, oxygen saturation \cite{koteska2022deep, shuzan2023machine}, respiratory rate \cite{shuzan2023machine, selvakumar2022realtime}, and blood glucose \cite{zhang2020noninvasive, jiang2025ppg}, as well as the detection and screening of conditions including arrhythmia \cite{pereira2020photoplethysmography, rantula2025photoplethysmography}, cardiac arrest \cite{shah2025automated, edgar2024automated}, hypertension \cite{elgendi2019use, min2025wearable}, and diabetes \cite{jiang2025ppg, avram2020digital}. Beyond these applications, PPG has also been explored in tasks such as emotion recognition \cite{ismail2022comparison, how2023towards}, stress analysis \cite{namvari2022photoplethysmography}, and individual identification \cite{zhang2023secure, wan2024deep}, motivated by inter-individual variability in waveform morphology and its association with autonomic nervous system regulation. More recently, advances in deep learning have further extended PPG-based research to cross-modal signal synthesis, such as the generation of ECG \cite{fang2025ppgflowecg, sarkar2021cardiogan, yuan2024catransformer, chen2025versatile} and arterial blood pressure \cite{bian2024constraint, chen2025versatile}, as well as the derivation of digital biomarkers related to cardiovascular health \cite{nie2025artificial, miller2025wearable}.

Despite the expanding range of PPG-based applications, most existing studies remain centered on a narrow set of cardiovascular-related tasks, leaving the broader potential of PPG for holistic health profiling largely underexplored.

\begin{figure*}
\centering
\includegraphics[width=0.8\linewidth]{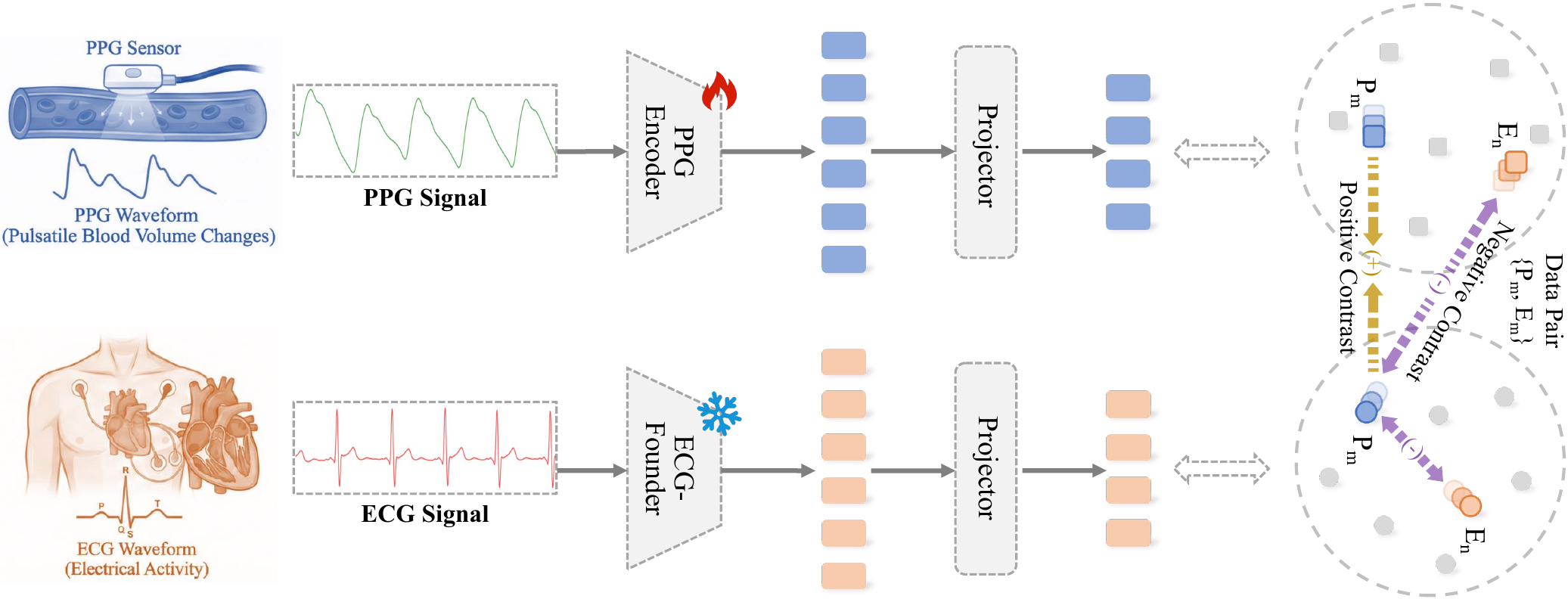}
\caption{ECG-guided cross-modal pretraining of AnyPPG. 
AnyPPG learns robust and physiologically grounded PPG representations via contrastive alignment guided by ECGFounder~\cite{li2025electrocardiogram}, an ECG foundation model, in a shared latent space, aligning synchronized PPG-ECG pairs while separating mismatched ones.}
\label{Figure1}
\end{figure*}

\section{Methodology}
In this section, we present AnyPPG, an ECG-guided PPG foundation model for investigating the extent to which PPG can support holistic health profiling. The use of ECG as a guiding modality is motivated by its intrinsic physiological coupling with PPG, where cardiac electrical activity governs peripheral blood volume dynamics while peripheral hemodynamics in turn modulate cardiac function, together with its demonstrated ability to encode rich information for both cardiovascular and multi-organ health \cite{li2026anyecg, hughes2025deep, fang2026ecgflowcmr, zhang2026ecgomics}. By aligning PPG representations with ECG, the model captures physiologically meaningful dynamics while suppressing non-physiological noise.

\subsection{Problem Formulation}
Let $\mathcal{D} = \{(x_{p,i}, x_{e,i})\}_{i=1}^{N}$ denote a dataset of $N$ synchronized PPG-ECG signal pairs, where $x_{p,i} \in \mathbb{R}^{L_p}$ and $x_{e,i} \in \mathbb{R}^{L_e}$ represent PPG and ECG segments of equal temporal duration. 
Our objective is to learn a parameterized PPG encoder $f_{\theta_p}(\cdot)$ that maps a raw PPG signal $x_{p,i}$ to a compact representation $z_{p,i} \in \mathbb{R}^{d_p}$, capturing transferable physiological dynamics for downstream tasks.

To this end, we introduce an auxiliary ECG encoder $f_{\theta_e}(\cdot)$ during pretraining and aim to align the representations of synchronized PPG-ECG pairs while separating mismatched pairs in the latent space. In this process, the ECG modality provides physiological supervision that guides the PPG encoder to capture hemodynamics consistent with cardiac electrical activity, thereby yielding robust and physiologically grounded PPG representations.

\subsection{Pretraining Framework}
To realize the cross-modal physiological alignment described above, AnyPPG adopts a dual-branch architecture composed of a PPG stream and an ECG stream. 
Each stream contains a modality-specific encoder followed by a projection head, enabling alignment of learned representations within a shared latent space~\cite{yin2024multi}. 
The overall framework is illustrated in Figure~\ref{Figure1}.

\paragraph{{\bfseries Modality-Specific Encoders}}
Each branch employs an encoder tailored to its input modality to extract latent physiological representations from raw waveforms. 
Given a synchronized pair $(x_{p,i}, x_{e,i})$, the encoders produce modality-specific latent features
\begin{equation}
z_{p,i} = f_{\theta_p}(x_{p,i}), \quad
z_{e,i} = f_{\theta_e}(x_{e,i}),
\end{equation}
where $z_{p,i}$ and $z_{e,i}$ denote the encoder-level representations of PPG and ECG signals, capturing complementary aspects of peripheral hemodynamics and cardiac electrical activity, respectively.

\paragraph{{\bfseries Projection Heads and Shared Latent Space}}
To enable cross-modal comparison, the encoder features are further mapped into a shared latent space through non-linear projection heads:
\begin{equation}
h_{p,i} = g_{\theta_p}(z_{p,i}), \quad
h_{e,i} = g_{\theta_e}(z_{e,i}).
\end{equation}
Here, $g_{\theta_p}(\cdot)$ and $g_{\theta_e}(\cdot)$ denote the projection heads (projectors) for the PPG and ECG branches, respectively, implemented as lightweight neural mappings placed on top of the encoders. 
The projected embeddings are further $\ell_2$-normalized to ensure scale-invariant similarity computation and stable contrastive optimization:
\begin{equation}
\hat{h}_{p,i} = \frac{h_{p,i}}{\|h_{p,i}\|_2}, \quad
\hat{h}_{e,i} = \frac{h_{e,i}}{\|h_{e,i}\|_2}.
\end{equation}

\paragraph{{\bfseries Cross-Modal Physiological Alignment}}
To leverage the intrinsic physiological coupling between cardiac electrical activity and peripheral hemodynamics, we adopt a CLIP-style contrastive learning objective~\cite{radford2021learning} for cross-modal alignment. Given a mini-batch of $B$ synchronized PPG-ECG pairs, the objective promotes correct cross-modal matching through a symmetric InfoNCE loss. 
Let $\mathrm{sim}(u, v) = u^{\top}v$ denote the similarity between two $\ell_2$-normalized vectors. 
The PPG-to-ECG alignment loss is defined as
\begin{equation}
\mathcal{L}_{p\to e} = -\frac{1}{B}\sum_{i=1}^{B}\log
\frac{\exp(\mathrm{sim}(h_{p,i}, h_{e,i}) / \tau)}
{\sum_{j=1}^{B}\exp(\mathrm{sim}(h_{p,i}, h_{e,j}) / \tau)},
\end{equation}
and the ECG-to-PPG alignment loss $\mathcal{L}_{e\to p}$ is vice verse. The overall training objective is
\begin{equation}
\mathcal{L} = \frac{1}{2}\left(\mathcal{L}_{p\to e} + \mathcal{L}_{e\to p}\right),
\end{equation}
where $\tau$ is a learnable temperature parameter that controls the sharpness of the similarity distribution. This contrastive objective increases similarity between synchronized PPG-ECG pairs while separating mismatched pairs, structuring the shared latent space according to physiologically consistent cardiovascular dynamics.

\begin{table*}
\small
\centering
\renewcommand\tabcolsep{11pt}
\renewcommand\arraystretch{1.0}
\caption{Datasets for pretraining and downstream evaluation. Task type: R (regression), B (binary), M-$k$ (multiclass with $k$ classes). Abbr.: HR, heart rate; SBP/DBP, systolic/diastolic blood pressure; RR, respiratory rate. $^{*}$ indicates complete exclusion from all pretraining splits.}
\begin{tabular}{lccccc}
\toprule
\textbf{Dataset} & \textbf{Used Modality} & \textbf{Task} & \textbf{Task Type} & \textbf{\#Subj. (Segments)} & \textbf{Recoding Hours} \\
\midrule

\multicolumn{6}{l}{\textbf{Phase I: Pretraining}} \\
MC-MED~\cite{kansal2025mc}   & PPG \& ECG &  & -- & 49,916 (28,420,140) & 78,945 \\
PulseDB~\cite{wang2023pulsedb}  & PPG \& ECG &  & -- & 4,964 (4,596,304)   & 12,768 \\
MESA~\cite{chen2015racial}     & PPG \& ECG & PPG-ECG alignment & -- & 2,010 (2,860,924) & 7,947 \\
HSP~\cite{sun2023human}      & PPG \& ECG &  & -- & 1,584 (3,333,705)   & 9,260 \\
CFS~\cite{redline1995familial}      & PPG \& ECG &  & -- & 322 (355,870)       & 989 \\
\textbf{Total} &  & & & \textbf{58,796 (39,566,943)} & \textbf{109,909} \\
\midrule

\multicolumn{6}{l}{\textbf{Phase II: Evaluation}} \\
PPG-DaLiA~\cite{reiss2019ppgdalia} & PPG & HR estimation  & R & 15 (12,943)        & 36 \\

          UCI-BP~\cite{kachuee2015cuffless}
          & PPG
          & SBP estimation & R & N/A (261,563)       & 727 \\
          &               & DBP estimation & R & N/A (261,563)   & 727 \\

          BUT PPG~\cite{nemcova2021brno}
          & PPG
          & HR estimation & R & 50 (3,840)         & 11 \\
          &               & SBP estimation & R & 50 (3,840)      & 11 \\
          &               & DBP estimation & R & 50 (3,840)      & 11 \\
          &               & Signal quality assessment & B & 50 (3,840) & 11 \\

          Gyro-Acc-PPG~\cite{lee2018motion} & PPG & HR estimation & R & 24 (2,016)        & 6 \\

          DeepBeat \cite{torres2020multi}
          & PPG
          &  Atrial fibrillation detection & B & N/A (536,399) & 1,490 \\

          WESAD \cite{schmidt2018introducing}
          & PPG
          &  Emotion Recognition & M-4 & 15 (4,419) & 12 \\

          Real-World PPG \cite{siam2019real}
          & PPG
          &  Biometric identification & B & 35 (20,74) & 3 \\ 

          WenXinWuYang \cite{wang2025reliability}
          & PPG
          & HR estimation & R & N/A (10,035) & 28 \\ 
          & PPG & RR estimation & R & N/A (10,035) & 28 \\ 
          & PPG & Age estimation & R & N/A (10,035) & 28 \\ 
          & PPG & Anomaly detection & B & N/A (10,035) & 28 \\ 
         
          MC-MED$^{*}$ \cite{kansal2025mc} & PPG & Phenome-wide disease detection & M-1468 & 15,759 (359,900) & 1,000 \\

\bottomrule
\end{tabular}
\label{Table1}
\end{table*}

\section{Experiments}
\subsection{Datasets and Splittings}
\paragraph{{\bfseries Pretraining Datasets for AnyPPG}}
As summarized in Table~\ref{Table1}, AnyPPG is pretrained on five large-scale publicly available datasets containing synchronized PPG and ECG recordings: MC-MED~\cite{kansal2025mc}, PulseDB~\cite{wang2023pulsedb}, the Multi-Ethnic Study of Atherosclerosis (MESA)~\cite{chen2015racial}, the Human Sleep Project (HSP)~\cite{sun2023human}, and the Cleveland Family Study (CFS)~\cite{redline1995familial}. PulseDB further aggregates waveform data from the MIMIC~\cite{johnson2016mimic} and VitalDB~\cite{lee2022vitaldb} databases. Detailed dataset descriptions are provided in Supplementary Section~A.1. Across these sources, we curate 109{,}909 hours of synchronized waveform recordings from 58{,}796 subjects, yielding approximately 40 million paired 10-second PPG-ECG segments. For pretraining monitoring and evaluation, the paired segments are partitioned at the subject level into training, validation, and test subsets using an 8:1:1 split.

\paragraph{{\bfseries Downstream Datasets for Evaluation}}
Following the proposed two-stage study paradigm, we first evaluate the representation quality of AnyPPG on diverse downstream datasets spanning both clinical and portable or wearable environments. 
As summarized in Table~\ref{Table1}, this evaluation includes eight PPG datasets: PPG-DaLiA~\cite{reiss2019ppgdalia}, UCI-BP~\cite{kachuee2015cuffless}, BUT PPG~\cite{nemcova2021brno}, Gyro-Acc-PPG~\cite{lee2018motion}, WESAD~\cite{schmidt2018introducing}, DeepBeat~\cite{torres2020multi}, Real-World PPG~\cite{siam2019real}, and WenXinWuYang~\cite{wang2025reliability}. 
Together, these datasets form fifteen dataset-task pairs covering a broad spectrum of physiological analysis tasks. Detailed dataset descriptions are provided in Supplementary Section~A.2. For data splitting, when official train-test splits are unavailable, we perform subject-level splitting with an 8:2 ratio for datasets containing identifiable subjects and random sample-level splitting otherwise. For datasets with predefined splits, we follow the original experimental protocols.

Building on this representation-level validation, we further employ AnyPPG as a tool for large-scale phenome-wide disease phenotype detection to assess the broader health-profiling potential of PPG signals. This analysis is conducted on the MC-MED dataset. Although MC-MED is included in the pretraining corpus, the phenome-wide evaluation results are reported on a strictly disjoint cohort consisting of patients excluded from all pretraining subsets (e.g., those without synchronized PPG-ECG recordings) and therefore never observed during pretraining, serving as a held-out test population independent of the pretraining training, validation, and test partitions. During evaluation, AnyPPG is fine-tuned using phenotype labels mapped from ICD-9/10 codes on the full set of subjects available in the pretraining cohort, and subsequently assessed on the strictly held-out cohort. For each hospitalization record, 20 PPG segments are randomly sampled from long-duration waveforms, and only phenotypes with more than 100 positive cases in the held-out test set are reported, resulting in a total of 1,468 phenotypes.

\subsection{Data Preprocessing}
For pretraining, continuous recordings are segmented into non-overlapping 10-second windows~\cite{orphanidou2017signal,koteska2022deep}. Segments containing more than 25\% invalid or motionless samples are discarded. The remaining segments are processed using band-pass filtering to suppress baseline drift and high-frequency noise: PPG signals are filtered within 0.5-8~Hz~\cite{elgendi2012analysis}, while ECG signals are filtered within 0.5-40~Hz and further denoised using a 50~Hz notch filter to remove powerline interference~\cite{li2025electrocardiogram}. ECG polarity inversion is automatically detected and corrected to ensure consistent waveform morphology~\cite{makowski2021neurokit2}, and ECG signal quality is assessed using established signal quality indices~\cite{zhao2018sqi}. In contrast, no explicit signal quality filtering is applied to PPG segments, allowing the model to learn robustness to noise and motion artifacts. All retained segments are resampled to a uniform sampling rate of 125~Hz for PPG and 500~Hz for ECG, followed by z-score normalization along the temporal dimension. For downstream evaluation, we follow the preprocessing protocols adopted in prior work~\cite{pillaipapagei}.

\subsection{Baselines and Implementation Details}
\paragraph{{ \bfseries Baseline Methods}} We benchmark AnyPPG against a set of representative baselines spanning three complementary categories: unimodal PPG self-supervised methods, general-purpose time-series foundation models, and PPG-specific foundation models. (i) Unimodal self-supervised learning: To evaluate the contribution of cross-modal physiological supervision in AnyPPG, we train SimCLR~\cite{chen2020simple} and BYOL~\cite{grill2020bootstrap} from scratch using only PPG signals from the same pretraining corpus; (ii) General-purpose time-series foundation models: We include MOMENT~\cite{goswami2024moment} and Chronos-2~\cite{ansari2025chronos} as representative general-purpose time-series foundation models. These models are evaluated as strong generic baselines to contextualize the performance of AnyPPG relative to broadly pretrained time-series representations; (iii) PPG-specific foundation models: PulsePPG~\cite{saha2025pulse} and PaPaGEI~\cite{pillaipapagei} are included as PPG-specific foundation models. These methods represent prior work most closely related to our setting and provide direct reference points for evaluating the performance of AnyPPG.

\paragraph{{ \bfseries Implementation Details}} For the implementation of AnyPPG, the PPG branch adopts Net1D~\cite{hong2020holmes}, a one-dimensional convolutional architecture derived from ResNet and widely used in prior PPG foundation models such as PaPaGEI~\cite{pillaipapagei} and PulsePPG~\cite{saha2025pulse}, producing a 512-dimensional encoder representation (see Supplementary Section~B for details). 
The ECG branch leverages ECGFounder~\cite{li2025electrocardiogram}, an ECG foundation model pretrained on over 10 million recordings with supervised cardiovascular diagnosis labels, whose encoder outputs a 1024-dimensional representation. Each projection head is implemented as a lightweight multilayer perceptron composed of two linear layers with Batch Normalization and GELU activation in between. During pretraining, the ECGFounder encoder is frozen, while its projection head is updated jointly with the PPG branch using the symmetric InfoNCE loss. The learnable temperature parameter $\tau$ is initialized to $0.07$. 

Pretraining is conducted on four NVIDIA H20 GPUs with a per-GPU batch size of 2{,}560. Optimization uses AdamW~\cite{loshchilov2017fixing} with an initial learning rate of $5\times10^{-4}$, a weight decay of $1\times10^{-2}$, and a cosine learning rate schedule. The model is trained for five epochs (77{,}160 optimization steps), including linear warm-up over the first 5{,}000 steps. Gradient clipping with a maximum norm of 1.0 is applied for training stability. Model checkpoints are saved every 500 steps, and the checkpoint achieving the lowest validation contrastive loss is selected for downstream evaluation. For SimCLR and BYOL baselines, we adopt identical pretraining settings to AnyPPG, including the same architecture, data, and optimization hyperparameters, ensuring a controlled comparison of learning objectives. For general-purpose time-series and PPG-specific foundation models, downstream evaluation is performed directly using their publicly released pretrained checkpoints.

\subsection{Performance Evaluation}
\paragraph{{\bfseries Evaluation Metrics}}
We evaluate both cross-modal representation alignment and downstream task performance using standard retrieval, regression, and classification metrics. For cross-modal alignment, retrieval quality is measured using Recall@k (R@1/5/10) and Mean Reciprocal Rank (MRR), capturing top-k matching accuracy and overall ranking consistency. All retrieval metrics are computed at the batch level and averaged across the evaluation set. For downstream tasks, Mean Absolute Error (MAE) is reported for regression, while AUC is used for classification. Additional metrics, including the Pearson correlation coefficient ($r$), accuracy, and F1-score, are provided in Supplementary Section~C for a more comprehensive evaluation. In multi-class settings, AUC is computed using a one-vs-rest strategy and macro-averaged across classes.

\paragraph{{\bfseries Linear Probing and Fine-Tuning Strategy}}
To assess representation quality on physiological analysis tasks, we adopt a linear probing protocol following prior work~\cite{pillaipapagei, saha2025pulse}. For all probing models, hyperparameters are selected via inner five-fold cross-validation on the training split, and final performance is reported on the held-out test set. For binary classification, logistic regression with the LBFGS solver is employed, with the inverse regularization strength searched over $C \in \{0.01, 0.1, 1, 10, 100\}$. For multi-class classification, we use a random forest classifier with grid-searched hyperparameters (number of estimators $\in \{100, 200\}$, maximum depth $\in \{10, 20\}$, and minimum samples split $\in \{2, 5\}$). For regression tasks, ridge regression is adopted, where the regularization parameter is chosen from $\alpha \in \{0.1, 1.0, 10.0, 100.0\}$ to minimize MAE. For phenome-wide disease phenotype detection, the full AnyPPG model is fine-tuned end-to-end from the pretrained checkpoint. Fine-tuning is formulated as a multi-label classification problem jointly optimized across all available disease phenotypes.

\section{Results}
In this section, we present the experimental results in three progressive stages. 
First, Section~\ref{result:pretrain_results} analyzes the cross-modal alignment achieved by the proposed pretraining strategy between PPG and ECG representations, demonstrating that AnyPPG learns physiologically meaningful PPG features consistent with ECG. 
Next, Section~\ref{result:downstream_results} evaluates the representation quality of AnyPPG across diverse downstream datasets, where it consistently achieves SOTA performance, confirming the robustness and generalizability of the learned representations. 
Finally, Section~\ref{result:phewas_results} extends the analysis to the large-scale phenome-wide setting, revealing broad associations between PPG signals and diverse disease phenotypes and highlighting the potential of PPG for holistic health profiling.

\subsection{AnyPPG Effectively Aligns PPG and ECG Representations in a Shared Latent Space}
\label{result:pretrain_results}
Table~\ref{Table2} summarizes PPG-to-ECG retrieval performance on the test splits of the five pretraining datasets. 
Overall, AnyPPG demonstrates strong and consistent cross-modal alignment, achieving sample-weighted Recall@1, Recall@5, and Recall@10 of 0.759, 0.946, and 0.970, respectively, with a MRR of 0.841. 
These results reflect both high retrieval accuracy and stable ranking consistency. 
Notably, performance remains robust across datasets with diverse acquisition settings and population characteristics, suggesting that the learned representations capture physiologically meaningful structures that generalize beyond individual data sources. 

\begin{table}
\centering
\renewcommand\tabcolsep{3pt}
\renewcommand\arraystretch{1.0}
\caption{PPG-to-ECG retrieval performance of AnyPPG across datasets. Metrics are computed using a fixed retrieval pool of 512 paired samples and averaged across all batches.}
\begin{tabular}{lrcccc}
\toprule
Dataset & \#Samples & R@1 & R@5 & R@10 & MRR \\
\midrule
MC-MED~\cite{kansal2025mc} & 2{,}796{,}347 & 0.786 & 0.947 & 0.966 & 0.857 \\
PulseDB~\cite{wang2023pulsedb} & 510{,}408 & 0.608 & 0.921 & 0.975 & 0.742 \\
MESA~\cite{chen2015racial} & 277{,}016 & 0.729 & 0.945 & 0.971 & 0.823 \\
HSP~\cite{sun2023human} & 333{,}059 & 0.796 & 0.978 & 0.994 & 0.876 \\
CFS~\cite{redline1995familial} & 35{,}854 & 0.644 & 0.910 & 0.957 & 0.760 \\
\midrule
Sample-weighted Avg. & \multirow{2}{*}{3{,}952{,}684} & 0.759 & 0.946& 0.970 & 0.841 \\
Macro Avg.       &                                    & 0.713 & 0.940 & 0.973 & 0.812 \\
\bottomrule
\end{tabular}
\label{Table2}
\end{table}

\subsection{AnyPPG Demonstrates Superior Performance Across Downstream Tasks}
\label{result:downstream_results}

\begin{table*}
\centering
\small
\setlength{\tabcolsep}{3pt} 

\newcommand{\CI}[3]{\ensuremath{#1_{\scriptscriptstyle[\,#2,\,#3\,]}}}

\newcommand{\best}[3]{\textcolor{red}{\textbf{\CI{#1}{#2}{#3}}}}
\newcommand{\second}[3]{\underline{\CI{#1}{#2}{#3}}}
\caption{Linear-probing MAE for regression tasks. Best results are in \textcolor{red}{red}, \underline{second best} are \underline{underlined}. $^{*}$ indicates identical pretraining data to AnyPPG. Values are reported with 95\% confidence intervals. Abbr.: HR, heart rate; SBP/DBP, systolic/diastolic blood pressure; RR, respiratory rate.}
\label{table:reg_result}

\resizebox{\linewidth}{!}{
\begin{tabular}{ll cccccccc}
\toprule
\textbf{Dataset} & \textbf{Task} &  
\makecell[c]{SimCLR$^{*}$ \\ ICML'20} &  
\makecell[c]{BYOL$^{*}$ \\ NIPS'20} &  
\makecell[c]{MOMENT \\ ICML'24} &  
\makecell[c]{Chronos-2 \\ ArXiv'25} &  
\makecell[c]{PulsePPG \\ IMWUT'25} &  
\makecell[c]{PaPaGEI-S \\ ICLR'25} &  
\makecell[c]{PaPaGEI-P \\ ICLR'25} &  
\makecell[c]{\textbf{AnyPPG} \\ \textbf{(Ours)}} \\
\midrule

\textbf{PPG-DaLiA} & HR &  
\second{9.37}{9.01}{9.74} & \CI{9.51}{9.17}{9.87} & \CI{10.43}{10.05}{10.79} & \CI{10.50}{10.14}{10.86} & \CI{10.79}{10.41}{11.20} & \CI{12.88}{12.50}{13.25} & \CI{11.69}{11.31}{12.04} & \best{9.27}{8.90}{9.62} \\

\midrule
\multirow{3}{*}{\textbf{BUT PPG}} & HR &  
\CI{7.63}{6.13}{9.19} & \CI{9.23}{7.52}{11.24} & \CI{7.35}{6.00}{8.88} & \second{7.05}{5.92}{8.27} & \CI{7.55}{5.98}{9.07} & \CI{8.92}{7.31}{10.51} & \CI{11.16}{8.28}{14.58} & \best{5.47}{4.39}{6.72} \\
& SBP &  
\CI{17.15}{14.23}{20.28} & \CI{17.71}{14.73}{20.73} & \CI{14.71}{12.07}{17.51} & \CI{14.55}{12.40}{17.09} & \CI{16.87}{14.41}{19.64} & \second{14.50}{11.59}{17.58} & \CI{25.76}{17.06}{36.51} & \best{13.31}{11.03}{15.68} \\
& DBP &  
\CI{11.15}{9.94}{12.51} & \CI{12.47}{10.91}{14.28} & \CI{10.75}{9.46}{12.05} & \CI{10.24}{9.15}{11.35} & \CI{11.32}{10.00}{12.77} & \CI{11.02}{9.73}{12.24} & \best{9.30}{7.92}{10.58} & \second{9.62}{8.41}{10.79} \\

\midrule
\textbf{Gyro-Acc-PPG} & HR &  
\second{11.38}{10.32}{12.55} & \CI{11.45}{10.47}{12.55} & \CI{13.34}{12.09}{14.57} & \CI{13.10}{12.16}{14.04} & \CI{12.56}{11.27}{13.72} & \CI{17.20}{15.84}{18.50} & \CI{14.63}{13.46}{15.76} & \best{10.82}{9.78}{11.85} \\

\midrule
\multirow{2}{*}{\textbf{UCI-BP}} & SBP &  
\second{15.70}{15.59}{15.80} & \CI{15.95}{15.84}{16.06} & \CI{16.57}{16.46}{16.68} & \CI{17.05}{16.94}{17.16} & \CI{16.54}{16.44}{16.65} & \CI{17.56}{17.45}{17.68} & \CI{16.80}{16.69}{16.92} & \best{15.62}{15.51}{15.72} \\
& DBP &  
\best{7.14}{7.09}{7.19} & \second{7.31}{7.26}{7.37} & \CI{7.53}{7.47}{7.58} & \CI{7.69}{7.63}{7.74} & \CI{7.57}{7.51}{7.63} & \CI{7.88}{7.82}{7.93} & \CI{7.68}{7.63}{7.73} & \best{7.14}{7.09}{7.19} \\

\midrule
\multirow{3}{*}{\textbf{WenXinWuYang}} & HR &  
\CI{7.14}{6.81}{7.47} & \second{7.07}{6.76}{7.39} & \CI{7.14}{6.80}{7.48} & \CI{7.22}{6.92}{7.55} & \CI{7.28}{6.95}{7.66} & \CI{9.47}{9.13}{9.82} & \CI{7.99}{7.63}{8.35} & \best{6.83}{6.50}{7.18} \\
& RR &  
\CI{2.33}{2.24}{2.42} & \second{2.32}{2.24}{2.40} & \CI{2.36}{2.24}{2.42} & \second{2.32}{2.23}{2.40} & \CI{2.36}{2.27}{2.44} & \CI{2.33}{2.24}{2.42} & \CI{2.33}{2.25}{2.41} & \best{2.30}{2.22}{2.39} \\
& Age &  
\second{6.09}{5.91}{6.29} & \best{6.08}{5.90}{6.27} & \CI{6.19}{5.99}{6.37} & \CI{6.19}{6.01}{6.37} & \CI{6.27}{6.05}{6.46} & \CI{6.47}{6.29}{6.67} & \CI{6.33}{6.15}{6.55} & \best{6.08}{5.89}{6.26} \\

\bottomrule
\end{tabular}}
\end{table*}

\begin{table*}
\centering
\small
\setlength{\tabcolsep}{3pt} 

\newcommand{\CI}[3]{\ensuremath{#1_{\scriptscriptstyle[\,#2,\,#3\,]}}}

\newcommand{\best}[3]{\textcolor{red}{\textbf{\CI{#1}{#2}{#3}}}}

\newcommand{\second}[3]{\underline{\CI{#1}{#2}{#3}}}
\caption{Linear-probing AUC for classification tasks. \textbf{Best} results are in \textcolor{red}{red}, \underline{second best} are underlined. $^{*}$ indicates identical pretraining data to AnyPPG. Values are reported with 95\% confidence intervals.}
\label{table:cls_result}
\resizebox{\linewidth}{!}{
\begin{tabular}{ll cccccccc}
\toprule
\textbf{Dataset} & \textbf{Task} &  
\makecell[c]{SimCLR$^{*}$ \\ ICML'20} &  
\makecell[c]{BYOL$^{*}$ \\ NIPS'20} &  
\makecell[c]{MOMENT \\ ICML'24} &  
\makecell[c]{Chronos-2 \\ ArXiv'25} &  
\makecell[c]{PulsePPG \\ IMWUT'25} &  
\makecell[c]{PaPaGEI-S \\ ICLR'25} &  
\makecell[c]{PaPaGEI-P \\ ICLR'25} &  
\makecell[c]{\textbf{AnyPPG} \\ \textbf{(Ours)}} \\
\midrule

\textbf{WESAD} & \makecell[l]{Emotion \\ recognition} &  
\best{0.77}{0.75}{0.79} & \CI{0.74}{0.72}{0.76} & \CI{0.71}{0.69}{0.73} & \CI{0.74}{0.72}{0.76} & \CI{0.73}{0.71}{0.75} & \CI{0.67}{0.65}{0.69} & \CI{0.71}{0.69}{0.73} & \second{0.76}{0.74}{0.78} \\

\midrule
\textbf{DeepBeat} & \makecell[l]{Atrial fibrillation  \\ detection} &  
\second{0.68}{0.67}{0.69} & \CI{0.64}{0.63}{0.65} & \CI{0.60}{0.59}{0.61} & \CI{0.66}{0.65}{0.67} & \CI{0.60}{0.59}{0.60} & \CI{0.57}{0.56}{0.58} & \CI{0.59}{0.58}{0.59} & \best{0.69}{0.68}{0.70} \\

\midrule
\textbf{BUT PPG} & \makecell[l]{Signal quality \\ assessment} &  
\second{0.73}{0.66}{0.79} & \best{0.79}{0.74}{0.84} & \CI{0.77}{0.72}{0.81} & \best{0.79}{0.74}{0.84} & \CI{0.76}{0.71}{0.81} & \CI{0.67}{0.61}{0.73} & \CI{0.69}{0.62}{0.76} & \best{0.79}{0.73}{0.84} \\

\midrule
\textbf{Real-World PPG} & \makecell[l]{Biometric \\ identification} &  
\best{1.00}{1.00}{1.00} & \best{1.00}{1.00}{1.00} & \CI{0.96}{0.96}{0.97} & \CI{0.88}{0.87}{0.90} & \best{1.00}{1.00}{1.00} & \CI{0.88}{0.87}{0.89} & \second{0.99}{0.99}{1.00} & \best{1.00}{1.00}{1.00} \\

\midrule
\textbf{WenXinWuYang} & \makecell[l]{Anomaly \\ detection} & 
\CI{0.67}{0.64}{0.70} & \CI{0.67}{0.64}{0.70} & \CI{0.68}{0.65}{0.71} & \second{0.69}{0.67}{0.72} & \second{0.69}{0.66}{0.72} & \CI{0.57}{0.55}{0.61} & \CI{0.68}{0.65}{0.71} & \best{0.72}{0.69}{0.75} \\
\bottomrule
\end{tabular}}
\end{table*}

Tables~\ref{table:reg_result} and~\ref{table:cls_result} summarize downstream performance for regression and classification tasks, respectively.
Overall, AnyPPG consistently outperforms all seven baseline models, achieving the best results on 13 of the 15 dataset-task pairs and ranking second on the remaining two, demonstrating strong and stable generalization across diverse downstream settings. Additional evaluation metrics and detailed results are provided in Supplementary Section~C.

Compared with unimodal self-supervised baselines (SimCLR and BYOL), AnyPPG yields notably larger improvements on datasets collected from real-world wearable and mobile devices, including BUT PPG and Gyro-Acc-PPG.
On these datasets, AnyPPG achieves average performance gains of approximately 15.4\% and 18.8\%, respectively.
Given that these data are characterized by motion artifacts and variable signal quality, the observed improvements suggest that cross-modal ECG supervision helps the model learn representations that are more robust under challenging acquisition conditions than those learned with unimodal self-supervised objectives.

When further compared with general-purpose time-series foundation models pretrained on large-scale, multi-domain temporal data (MOMENT and Chronos-2), AnyPPG achieves the best performance on all 15 tasks. While these models exhibit competitive results in specific areas, such as signal quality assessment on the BUT PPG dataset, AnyPPG consistently outperforms them across the board. This disparity suggests that generic temporal representations may fail to fully capture the intricate physiological characteristics inherent to PPG signals. In contrast, the physiologically grounded cross-modal alignment in AnyPPG facilitates more effective transfer to PPG-centric downstream applications.

Finally, compared with existing PPG-specific foundation models, AnyPPG achieves superior performance on 14 of the 15 tasks, underscoring the benefit of combining cross-modal physiological supervision with large-scale, multi-source pretraining. Notably, general self-supervised frameworks (SimCLR and BYOL), when trained on the same pretraining corpus as AnyPPG, also outperform PPG-specific models on a subset of tasks, suggesting that the scale and diversity of pretraining data play a critical role in determining representation quality. Collectively, these findings indicate that AnyPPG provides highly effective and generalizable representations for PPG analysis across diverse downstream settings.

\subsection{AnyPPG Reveals the Potential of PPG for Holistic Health Profiling}
\label{result:phewas_results}

\begin{figure*}
\centering
\includegraphics[width=\linewidth]{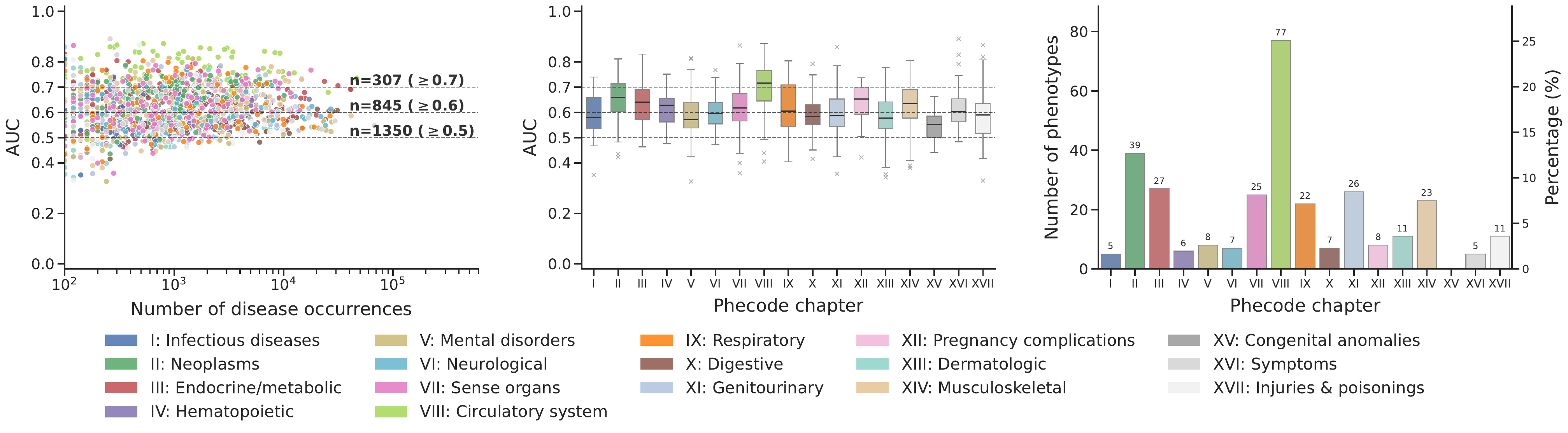}
\caption{Phenome-wide disease detection performance of AnyPPG. Left: Distribution of AUC scores for 1,468 disease phenotypes. Middle: Chapter-level AUC performance distribution. Right: Count and percentage distribution of high-performing phenotypes (AUC $\ge$ 0.70) across phecode chapters, with counts annotated above each bar.}
\label{fig:icd10_combined}
\end{figure*}

\begin{figure*}
    \centering
    \includegraphics[width=\linewidth]{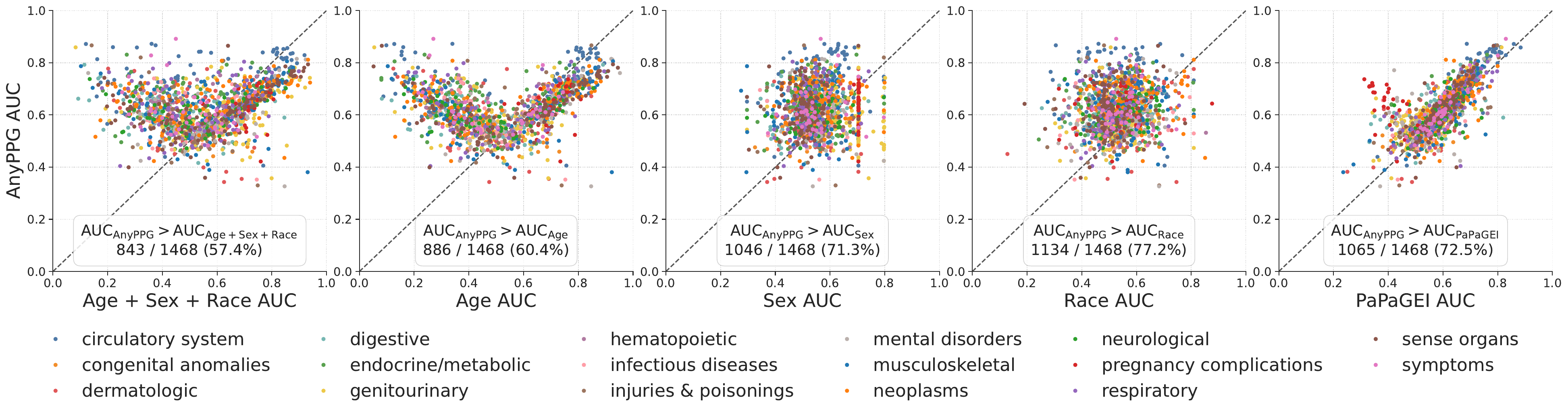}
    \caption{Phenotype-level AUC comparison with baseline models. Each point represents a disease phenotype. The dashed line indicates equal performance. Points above the diagonal correspond to phenotypes where AnyPPG outperforms the baseline.}
    \label{fig:baseline_comparison}
\end{figure*}

In the phenome-wide disease detection study comprising a total of 1,468 disease phenotypes, AnyPPG demonstrates discriminative capacity across a broad spectrum of diseases rather than being limited to cardiovascular conditions. This global discriminative capacity is first reflected at the phenotype level (left panel of Figure~\ref{fig:icd10_combined}), where 307 of the 1,468 evaluated phenotypes achieve an AUC of at least 0.70, and 845 and 1,350 phenotypes exceed AUC thresholds of 0.60 and 0.50, respectively, indicating that the vast majority of disease phenotypes exhibit detectable patterns with at least partial discriminative ability based solely on PPG signals. Consistent with this phenotype-level observation, chapter-level aggregation (middle panel) shows mean AUC values above the 0.50 baseline across all phecode chapters, with several non-circulatory categories, such as endocrine/metabolic disorders and hematopoietic diseases, surpassing 0.60. Importantly, the right panel reveals that high-performing phenotypes are not confined to circulatory diseases but are broadly distributed across organ systems. Chapters including neoplasms, endocrine/metabolic disorders, sense organs, respiratory, genitourinary, and musculoskeletal conditions each contain more than 20 phenotypes with AUC values above 0.70. To further contextualize these results, we compare AnyPPG with a set of demographic baselines constructed from age, sex, and race (and their combinations), as well as a PPG foundation model (PaPaGEI-S). As shown in Figure~\ref{fig:baseline_comparison}, AnyPPG outperforms the strongest demographic baseline (age, sex, and race) in 843 of 1,468 (57.4\%) phenotypes and surpasses PaPaGEI-S in 1,065 of 1,468 (72.5\%) phenotypes, indicating that the learned PPG representations provide additional discriminative information beyond both demographic factors and existing PPG foundation models.

To further characterize high-performing phenotypes, we examined the top 60 conditions ranked by AUC (see Supplementary Section~E, Figure~5 for the full list). All Top-60 phenotypes achieve AUC values of at least 0.78, indicating consistently strong discriminative performance. Circulatory diseases account for 31 of the 60 phenotypes and can be broadly grouped into three categories: (i) cardiac pump dysfunction, such as heart failure; (ii) structural and valvular abnormalities, including mitral or tricuspid valve disorders; and (iii) electrophysiological conduction disturbances, such as atrial fibrillation and left bundle branch block. Importantly, strong performance extends well beyond cardiovascular conditions. The remaining 29 phenotypes arise from ten non-circulatory chapters, demonstrating the broad physiological relevance of PPG-derived representations. Representative examples include diabetic retinopathy (AUC = 0.83) and amyloidosis (AUC = 0.81) within endocrine and metabolic disorders, dementia-related phenotypes (AUC = 0.81) in the neurological chapter, and corneal dystrophy (AUC = 0.86) and wet macular degeneration (AUC = 0.79) in sense organs, many of which have rarely been investigated using PPG in prior work. Moreover, the Top-60 results represent only a small portion of the total 307 discriminative phenotypes (detailed in Supplementary Section~E). Meaningful discriminative capability is also observed across numerous additional clinically meaningful non-circulatory conditions, such as Parkinson’s disease (AUC = 0.77), hyperkalemia (AUC = 0.75), chronic renal failure (AUC = 0.75), and anemia of chronic disease (AUC = 0.72).

Collectively, these findings suggest that PPG reflects integrated physiological alterations across a broad spectrum of conditions, such as metabolic, neurological, and hematological disorders \cite{duh2017diabetic, valenza2025brain}, supporting its potential for holistic health profiling.

\section{Limitations and Ethical Considerations}
This study has several limitations. First, the phenome-wide evaluation was conducted in a single clinical cohort, and the observed associations may therefore reflect cohort-specific characteristics. External validation across additional medical centers and more diverse populations is needed to assess robustness and generalizability. Moreover, although the phenome-wide analysis identifies disease phenotypes discriminable from PPG-derived representations, these findings should be interpreted as quantifying broad associations rather than establishing causal relationships.

All datasets used in this study, except for the WenXinWuYang dataset, are publicly available, and the requirements for Institutional Review Board approval and informed consent were waived accordingly. The proprietary WenXinWuYang dataset was collected using consumer-grade devices and contains only de-identified physiological signals and basic health-related metrics. Its use was reviewed and approved by the Peking University Medical Ethics Committee (Approval No.: IRB00001052-23071). 

\section{Conclusion}
In this work, we introduced AnyPPG, an ECG-guided PPG foundation model that achieves state-of-the-art performance in PPG analysis and enables investigation of the extent to which PPG can support holistic health profiling beyond traditional cardiovascular tasks. Through large-scale phenome-wide analysis on over 15,000 subjects, we show that PPG representations learned by AnyPPG capture clinically relevant information across both circulatory and non-circulatory conditions. These findings provide a foundation for future research exploring the broader role of PPG in holistic health assessment.

\section*{GenAI Disclosure}
During this research and manuscript preparation, we use large language models (LLMs) for auxiliary tasks such as generating short scripts during coding and assisting with text translation and language polishing. All core ideas, research methodologies, and academic contributions are conceived and developed independently by the authors, with the role of LLMs limited to improving the fluency and readability of the presentation.

\section*{Acknowledgements}
This work is supported by the National Natural Science Foundation of China (62102008, 62172018), CCF-Tencent Rhino-Bird Open Research Fund (CCF-Tencent RAGR20250108), CCF-Zhipu Large Model Innovation Fund (CCF-Zhipu202414), PKU-OPPO Fund (BO202301, BO202503), Research Project of Peking University in the State Key Laboratory of Vascular Homeostasis and Remodeling (2025-SKLVHR-YCTS-02), Beijing Municipal Science and Technology Commission (Z251100000725008), Prevention and Control of Emerging and Major Infectious Diseases-National Science and Technology Major Project (2025ZD01906000, 2025ZD01906004), Capital’s Funds for Health Improvement and Research (CFH2026-1-4092), Beijing Natural Science Foundation (QY26080). We also thank Liwei Liu and Guowei Li from Huawei Technologies Co., Ltd for valuable discussions.

\section*{Supplementary Materials}
Due to space constraints, supplementary materials (including complete dataset descriptions, detailed encoder architecture specifications, additional evaluation metrics, ablation experiments, subgroup analyses, and full phenome-wide disease detection results) are available at \url{https://github.com/PKUDigitalHealth/AnyPPG}.


\bibliographystyle{ACM-Reference-Format}
\bibliography{reference}

@article{bayoumy2021smart,
  title={Smart wearable devices in cardiovascular care: where we are and how to move forward},
  author={Bayoumy, Karim and Gaber, Mohammed and Elshafeey, Abdallah and Mhaimeed, Omar and Dineen, Elizabeth H and Marvel, Francoise A and Martin, Seth S and Muse, Evan D and Turakhia, Mintu P and Tarakji, Khaldoun G and others},
  journal={Nature Reviews Cardiology},
  volume={18},
  number={8},
  pages={581--599},
  year={2021},
  publisher={Nature Publishing Group UK London}
}

@article{williams2023wearable,
  title={Wearable technology and the cardiovascular system: the future of patient assessment},
  author={Williams, Gareth J and Al-Baraikan, Abdulaziz and Rademakers, Frank E and Ciravegna, Fabio and van de Vosse, Frans N and Lawrie, Allan and Rothman, Alexander and Ashley, Euan A and Wilkins, Martin R and Lawford, Patricia V and others},
  journal={The Lancet Digital Health},
  volume={5},
  number={7},
  pages={e467--e476},
  year={2023},
  publisher={Elsevier}
}

@article{spatz2024wearable,
  title={Wearable digital health technologies for monitoring in cardiovascular medicine},
  author={Spatz, Erica S and Ginsburg, Geoffrey S and Rumsfeld, John S and Turakhia, Mintu P},
  journal={New England Journal of Medicine},
  volume={390},
  number={4},
  pages={346--356},
  year={2024},
  publisher={Mass Medical Soc}
}

@article{jiang2025ppg,
  title={PPG-based glucose sensors: a review},
  author={Jiang, Hui and Yao, Tianliang and Ding, Cheng},
  journal={Artificial Intelligence Review},
  volume={58},
  number={12},
  pages={1--33},
  year={2025},
  publisher={Springer}
}

@article{nie2024review,
  title={A review of deep learning methods for photoplethysmography data},
  author={Nie, Guangkun and Zhu, Jiabao and Tang, Gongzheng and Zhang, Deyun and Geng, Shijia and Zhao, Qinghao and Hong, Shenda},
  journal={arXiv preprint arXiv:2401.12783},
  year={2024}
}

@article{bent2020investigating,
  title={Investigating sources of inaccuracy in wearable optical heart rate sensors},
  author={Bent, Brinnae and Goldstein, Benjamin A and Kibbe, Warren A and Dunn, Jessilyn P},
  journal={NPJ digital medicine},
  volume={3},
  number={1},
  pages={18},
  year={2020},
  publisher={Nature Publishing Group UK London}
}

@article{elgendi2019use,
  title={The use of photoplethysmography for assessing hypertension},
  author={Elgendi, Mohamed and Fletcher, Richard and Liang, Yongbo and Howard, Newton and Lovell, Nigel H and Abbott, Derek and Lim, Kenneth and Ward, Rabab},
  journal={NPJ digital medicine},
  volume={2},
  number={1},
  pages={60},
  year={2019},
  publisher={Nature Publishing Group UK London}
}

@article{pereira2020photoplethysmography,
  title={Photoplethysmography based atrial fibrillation detection: a review},
  author={Pereira, Tania and Tran, Nate and Gadhoumi, Kais and Pelter, Michele M and Do, Duc H and Lee, Randall J and Colorado, Rene and Meisel, Karl and Hu, Xiao},
  journal={NPJ digital medicine},
  volume={3},
  number={1},
  pages={3},
  year={2020},
  publisher={Nature Publishing Group UK London}
}

@article{shah2025automated,
  title={Automated loss of pulse detection on a consumer smartwatch},
  author={Shah, Kamal and Wang, Anran and Chen, Yiwen and Munjal, Jitender and Chhabra, Sumeet and Stange, Anthony and Wei, Enxun and Phan, Tuan and Giest, Tracy and Hawkins, Beszel and others},
  journal={Nature},
  pages={1--3},
  year={2025},
  publisher={Nature Publishing Group UK London}
}

@article{edgar2024automated,
  title={Automated cardiac arrest detection using a photoplethysmography wristband: algorithm development and validation in patients with induced circulatory arrest in the DETECT-1 study},
  author={Edgar, Roos and Scholte, Niels TB and Ebrahimkheil, Kambiz and Brouwer, Marc A and Beukema, Rypko J and Mafi-Rad, Masih and Vernooy, Kevin and Yap, Sing-Chien and Ronner, Eelko and van Mieghem, Nicolas and others},
  journal={The Lancet Digital Health},
  volume={6},
  number={3},
  pages={e201--e210},
  year={2024},
  publisher={Elsevier}
}

@article{nie2025artificial,
  title={Artificial Intelligence-derived Vascular Age from Photoplethysmography: A Novel Digital Biomarker for Cardiovascular Health},
  author={Nie, Guangkun and Zhao, Qinghao and Tang, Gongzheng and Li, Yaxin and Hong, Shenda},
  journal={arXiv preprint arXiv:2502.12990},
  year={2025}
}

@article{miller2025wearable,
  title={A wearable-based aging clock associates with disease and behavior},
  author={Miller, Andrew C and Futoma, Joseph and Abbaspourazad, Salar and Heinze-Deml, Christina and Emrani, Saba and Shapiro, Ian and Sapiro, Guillermo},
  journal={Nature communications},
  volume={16},
  number={1},
  pages={9264},
  year={2025},
  publisher={Nature Publishing Group UK London}
}

@article{holmstrom2023deep,
  title={Deep learning-based electrocardiographic screening for chronic kidney disease},
  author={Holmstrom, Lauri and Christensen, Matthew and Yuan, Neal and Weston Hughes, J and Theurer, John and Jujjavarapu, Melvin and Fatehi, Pedram and Kwan, Alan and Sandhu, Roopinder K and Ebinger, Joseph and others},
  journal={Communications Medicine},
  volume={3},
  number={1},
  pages={73},
  year={2023},
  publisher={Nature Publishing Group UK London}
}

@article{lin2026ai,
  title={AI-enabled electrocardiogram alert for potassium imbalance treatment: a pragmatic randomized controlled trial},
  author={Lin, Chin and Lin, Chin-Sheng and Chen, Sy-Jou and Tsai, Shi-Hung and Sung, Chih-Chien and Chen, Chien-Chou and Hsu, Yu-Juei and Hung, Yi-Jen and Lin, Shih-Hua},
  journal={Nature Communications},
  volume={17},
  number={1},
  pages={159},
  year={2026},
  publisher={Nature Publishing Group UK London}
}

@article{hughes2025deep,
  title={A deep learning phenome wide association study of the electrocardiogram},
  author={Hughes, John Weston and Theurer, John and Vukadinovic, Milos and Rogers, Albert J and Somani, Sulaiman and Kang, Guson and Ghazizadeh, Zaniar and O’Sullivan, Jack W and Jain, Sneha S and Gomes, Bruna and others},
  journal={European Heart Journal-Digital Health},
  pages={ztaf047},
  year={2025},
  publisher={Oxford University Press UK}
}

@article{li2026anyecg,
  title={AnyECG: Evolved ECG Foundation Model for Holistic Health Profiling},
  author={Li, Jun and Zhu, Hongling and Xiao, Yujie and Zhao, Qinghao and Ke, Yalei and Tang, Gongzheng and Nie, Guangkun and Zhang, Deyun and Li, Jin and Yu, Canqing and others},
  journal={arXiv preprint arXiv:2601.10748},
  year={2026}
}

@article{alcaraz2025electrocardiogram,
  title={Electrocardiogram-based diagnosis of liver diseases: an externally validated and explainable machine learning approach},
  author={Alcaraz, Juan Miguel Lopez and Haverkamp, Wilhelm and Strodthoff, Nils},
  journal={EClinicalMedicine},
  volume={84},
  year={2025},
  publisher={Elsevier}
}

@article{simonetto2025detection,
  title={Detection of undiagnosed liver cirrhosis via AI-enabled electrocardiogram: a pragmatic, cluster-randomized clinical trial},
  author={Simonetto, Douglas A and Rushlow, David and Liu, Kan and Calleri, Alberto and Kassmeyer, Blake A and Lennon, Ryan J and Rattan, Puru and Bernard, Matthew E and Singh, Gagandeep and Deyo-Svendsen, Mark E and others},
  journal={Nature Medicine},
  pages={1--8},
  year={2025},
  publisher={Nature Publishing Group US New York}
}

@article{abbaspourazad2023large,
  title={Large-scale training of foundation models for wearable biosignals},
  author={Abbaspourazad, Salar and Elachqar, Oussama and Miller, Andrew C and Emrani, Saba and Nallasamy, Udhyakumar and Shapiro, Ian},
  journal={arXiv preprint arXiv:2312.05409},
  year={2023}
}

@article{zhang2020noninvasive,
  title={A noninvasive blood glucose monitoring system based on smartphone PPG signal processing and machine learning},
  author={Zhang, Gaobo and Mei, Zhen and Zhang, Yuan and Ma, Xuesheng and Lo, Benny and Chen, Dongyi and Zhang, Yuanting},
  journal={IEEE Transactions on Industrial Informatics},
  volume={16},
  number={11},
  pages={7209--7218},
  year={2020},
  publisher={IEEE}
}

@article{rantula2025photoplethysmography,
  title={Photoplethysmography in recent-onset atrial fibrillation: automatic detection of rhythm change and burden},
  author={Rantula, Olli A and Lipponen, Jukka A and Halonen, Jari and J{\"a}ntti, Helena and Rissanen, Tuomas T and Naukkarinen, Noora S and V{\"a}liaho, Eemu-Samuli and Santala, Onni E and Sedha, Jagdeep and Martikainen, Tero J and others},
  journal={European Heart Journal-Digital Health},
  pages={ztaf055},
  year={2025},
  publisher={Oxford University Press UK}
}

@article{min2025wearable,
  title={Wearable blood pressure sensors for cardiovascular monitoring and machine learning algorithms for blood pressure estimation},
  author={Min, Seongwook and An, Jaehun and Lee, Jae Hee and Kim, Ji Hoon and Joe, Daniel J and Eom, Soo Hwan and Yoo, Chang D and Ahn, Hyo-Suk and Hwang, Jin-Young and Xu, Sheng and others},
  journal={Nature Reviews Cardiology},
  pages={1--20},
  year={2025},
  publisher={Nature Publishing Group UK London}
}

@article{avram2020digital,
  title={A digital biomarker of diabetes from smartphone-based vascular signals},
  author={Avram, Robert and Olgin, Jeffrey E and Kuhar, Peter and Hughes, J Weston and Marcus, Gregory M and Pletcher, Mark J and Aschbacher, Kirstin and Tison, Geoffrey H},
  journal={Nature medicine},
  volume={26},
  number={10},
  pages={1576--1582},
  year={2020},
  publisher={Nature Publishing Group US New York}
}

@article{ismail2022comparison,
  title={A comparison of emotion recognition system using electrocardiogram (ECG) and photoplethysmogram (PPG)},
  author={Ismail, Sharifah Noor Masidayu Sayed and Aziz, Nor Azlina Ab and Ibrahim, Siti Zainab},
  journal={Journal of King Saud University-Computer and Information Sciences},
  volume={34},
  number={6},
  pages={3539--3558},
  year={2022},
  publisher={Elsevier}
}

@article{how2023towards,
  title={Towards PPG-based anger detection for emotion regulation},
  author={How, Tuck-Voon and Green, Robin EA and Mihailidis, Alex},
  journal={Journal of NeuroEngineering and Rehabilitation},
  volume={20},
  number={1},
  pages={107},
  year={2023},
  publisher={Springer}
}

@article{namvari2022photoplethysmography,
  title={Photoplethysmography enabled wearable devices and stress detection: a scoping review},
  author={Namvari, Mina and Lipoth, Jessica and Knight, Sheida and Jamali, Ali Akbar and Hedayati, Mojtaba and Spiteri, Raymond J and Syed-Abdul, Shabbir},
  journal={Journal of Personalized Medicine},
  volume={12},
  number={11},
  pages={1792},
  year={2022},
  publisher={MDPI}
}

@article{zhang2023secure,
  title={A secure, flexible, and PPG-based biometric scheme for healthy IoT using homomorphic random forest},
  author={Zhang, Liping and Li, Anzi and Chen, Shukai and Ren, Wei and Choo, Kim-Kwang Raymond},
  journal={IEEE Internet of Things Journal},
  volume={11},
  number={1},
  pages={612--622},
  year={2023},
  publisher={IEEE}
}

@article{wan2024deep,
  title={Deep learning-based photoplethysmography biometric authentication for continuous user verification},
  author={Wan, Li and Liu, Kechen and Mengash, Hanan Abdullah and Alruwais, Nuha and Al Duhayyim, Mesfer and Venkatachalam, K},
  journal={Applied Soft Computing},
  volume={156},
  pages={111461},
  year={2024},
  publisher={Elsevier}
}

@inproceedings{sarkar2021cardiogan,
  title={Cardiogan: Attentive generative adversarial network with dual discriminators for synthesis of ecg from ppg},
  author={Sarkar, Pritam and Etemad, Ali},
  booktitle={Proceedings of the AAAI Conference on Artificial Intelligence},
  volume={35},
  number={1},
  pages={488--496},
  year={2021}
}

@article{yuan2024catransformer,
  title={CATransformer: A cycle-aware transformer for high-fidelity ECG generation from PPG},
  author={Yuan, Xiaoyan and Wang, Wei and Li, Xiaohe and Zhang, Yuanting and Hu, Xiping and Deen, M Jamal},
  journal={IEEE Journal of Biomedical and Health Informatics},
  year={2024},
  publisher={IEEE}
}

@article{chen2025versatile,
  title={Versatile cardiovascular signal generation with a unified diffusion transformer},
  author={Chen, Zehua and Miao, Yuyang and Wang, Liyuan and Fan, Luyun and Mandic, Danilo P and Zhu, Jun},
  journal={Nature Machine Intelligence},
  pages={1--14},
  year={2025},
  publisher={Nature Publishing Group UK London}
}

@article{saha2025pulse,
  title={Pulse-ppg: An open-source field-trained ppg foundation model for wearable applications across lab and field settings},
  author={Saha, Mithun and Xu, Maxwell A and Mao, Wanting and Neupane, Sameer and Rehg, James M and Kumar, Santosh},
  journal={Proceedings of the ACM on Interactive, Mobile, Wearable and Ubiquitous Technologies},
  volume={9},
  number={3},
  pages={1--35},
  year={2025},
  publisher={ACM New York, NY, USA}
}

@article{chen2025gpt,
  title={GPT-PPG: a GPT-based foundation model for photoplethysmography signals},
  author={Chen, Zhaoliang and Ding, Cheng and Kataria, Saurabh and Yan, Runze and Wang, Minxiao and Lee, Randall and Hu, Xiao},
  journal={Physiological Measurement},
  volume={46},
  number={5},
  pages={055004},
  year={2025},
  publisher={IOP Publishing}
}

@article{ding2024siamquality,
  title={SiamQuality: a ConvNet-based foundation model for photoplethysmography signals},
  author={Ding, Cheng and Guo, Zhicheng and Chen, Zhaoliang and Lee, Randall J and Rudin, Cynthia and Hu, Xiao},
  journal={Physiological Measurement},
  volume={45},
  number={8},
  pages={085004},
  year={2024},
  publisher={IOP Publishing}
}

@article{xu2024whole,
  title={A whole-slide foundation model for digital pathology from real-world data},
  author={Xu, Hanwen and Usuyama, Naoto and Bagga, Jaspreet and Zhang, Sheng and Rao, Rajesh and Naumann, Tristan and Wong, Cliff and Gero, Zelalem and Gonz{\'a}lez, Javier and Gu, Yu and others},
  journal={Nature},
  volume={630},
  number={8015},
  pages={181--188},
  year={2024},
  publisher={Nature Publishing Group UK London}
}

@article{lu2024multimodal,
  title={A multimodal generative AI copilot for human pathology},
  author={Lu, Ming Y and Chen, Bowen and Williamson, Drew FK and Chen, Richard J and Zhao, Melissa and Chow, Aaron K and Ikemura, Kenji and Kim, Ahrong and Pouli, Dimitra and Patel, Ankush and others},
  journal={Nature},
  volume={634},
  number={8033},
  pages={466--473},
  year={2024},
  publisher={Nature Publishing Group UK London}
}

@article{ma2025fully,
  title={A fully open AI foundation model applied to chest radiography},
  author={Ma, DongAo and Pang, Jiaxuan and Gotway, Michael B and Liang, Jianming},
  journal={Nature},
  pages={1--11},
  year={2025},
  publisher={Nature Publishing Group UK London}
}

@article{sun2025foundation,
  title={A foundation model for enhancing magnetic resonance images and downstream segmentation, registration and diagnostic tasks},
  author={Sun, Yue and Wang, Limei and Li, Gang and Lin, Weili and Wang, Li},
  journal={Nature Biomedical Engineering},
  volume={9},
  number={4},
  pages={521--538},
  year={2025},
  publisher={Nature Publishing Group UK London}
}

@article{li2025electrocardiogram,
  title={An Electrocardiogram Foundation Model Built on over 10 Million Recordings},
  author={Li, Jun and Aguirre, Aaron D and Junior, Valdery Moura and Jin, Jiarui and Liu, Che and Zhong, Lanhai and Sun, Chenxi and Clifford, Gari and Brandon Westover, M and Hong, Shenda},
  journal={NEJM AI},
  volume={2},
  number={7},
  pages={AIoa2401033},
  year={2025},
  publisher={Massachusetts Medical Society}
}

@article{thapa2026multimodal,
  title={A multimodal sleep foundation model for disease prediction},
  author={Thapa, Rahul and Kjaer, Magnus Ruud and He, Bryan and Covert, Ian and Moore IV, Hyatt and Hanif, Umaer and Ganjoo, Gauri and Westover, M Brandon and Jennum, Poul and Brink-Kjaer, Andreas and others},
  journal={Nature Medicine},
  pages={1--11},
  year={2026},
  publisher={Nature Publishing Group US New York}
}

@article{zhang2025sensorlm,
  title={SensorLM: Learning the Language of Wearable Sensors},
  author={Zhang, Yuwei and Ayush, Kumar and Qiao, Siyuan and Heydari, A Ali and Narayanswamy, Girish and Xu, Maxwell A and Metwally, Ahmed A and Xu, Shawn and Garrison, Jake and Xu, Xuhai and others},
  journal={arXiv preprint arXiv:2506.09108},
  year={2025}
}

@article{guo2023siamaf,
  title={SiamAF: learning shared information from ECG and PPG signals for robust atrial fibrillation detection},
  author={Guo, Zhicheng and Ding, Cheng and Do, Duc H and Shah, Amit and Lee, Randall J and Hu, Xiao and Rudin, Cynthia},
  journal={arXiv preprint arXiv:2310.09203},
  year={2023}
}

@inproceedings{bian2024constraint,
  title={Constraint latent space matters: an anti-anomalous waveform transformation solution from photoplethysmography to arterial blood pressure},
  author={Bian, Cheng and Li, Xiaoyu and Bi, Qi and Zhu, Guangpu and Lyu, Jiegeng and Zhang, Weile and Li, Yelei and Zeng, Zijing},
  booktitle={Proceedings of the AAAI Conference on Artificial Intelligence},
  volume={38},
  number={10},
  pages={11087--11095},
  year={2024}
}

@inproceedings{schlesinger2020blood,
  title={Blood pressure estimation from ppg signals using convolutional neural networks and siamese network},
  author={Schlesinger, Oded and Vigderhouse, Nitai and Eytan, Danny and Moshe, Yair},
  booktitle={ICASSP 2020-2020 IEEE international conference on acoustics, speech and signal processing (ICASSP)},
  pages={1135--1139},
  year={2020},
  organization={IEEE}
}

@article{panwar2020pp,
  title={PP-Net: A deep learning framework for PPG-based blood pressure and heart rate estimation},
  author={Panwar, Madhuri and Gautam, Arvind and Biswas, Dwaipayan and Acharyya, Amit},
  journal={IEEE Sensors Journal},
  volume={20},
  number={17},
  pages={10000--10011},
  year={2020},
  publisher={IEEE}
}

@article{nie2025low,
  title={A Low-Burden Sleep Foundation Model Built on Respiratory and Heartbeat Signals from 780,000+ Hours of Multi-Ethnic Sleep Recordings},
  author={Nie, Guangkun and Chen, Xuesong and Wang, Yichen and Chen, Jingxu and Shi, Yunhan and Zhong, Jianwen and Huang, Weijun and Jin, Zengrui and Lei, Fei and Wang, Leilei and others},
  journal={medRxiv},
  pages={2025--09},
  year={2025},
  publisher={Cold Spring Harbor Laboratory Press}
}

@inproceedings{koteska2022deep,
  title={A deep learning approach to estimate SpO2 from PPG signals},
  author={Koteska, Bojana and Bodanova, Ana Madevska and Mitrova, Hristina and Sidorenko, Marija and Lehocki, Fedor},
  booktitle={Proceedings of the 9th International Conference on Bioinformatics Research and Applications},
  pages={142--148},
  year={2022}
}

@article{shuzan2023machine,
  title={Machine learning-based respiration rate and blood oxygen saturation estimation using photoplethysmogram signals},
  author={Shuzan, Md Nazmul Islam and Chowdhury, Moajjem Hossain and Chowdhury, Muhammad EH and Murugappan, Murugappan and Hoque Bhuiyan, Enamul and Arslane Ayari, Mohamed and Khandakar, Amith},
  journal={Bioengineering},
  volume={10},
  number={2},
  pages={167},
  year={2023},
  publisher={MDPI}
}

@article{selvakumar2022realtime,
  title={Realtime PPG based respiration rate estimation for remote health monitoring applications},
  author={Selvakumar, K and Kumar, E Vinodh and Sailesh, M and Varun, Mamtani and Allan, Anbu and Biswajit, Nanda and Namrata, Panga and Upasana, Sivaramakrishnan},
  journal={Biomedical Signal Processing and Control},
  volume={77},
  pages={103746},
  year={2022},
  publisher={Elsevier}
}

@article{fang2025ppgflowecg,
  title={PPGFlowECG: Latent Rectified Flow with Cross-Modal Encoding for PPG-Guided ECG Generation and Cardiovascular Disease Detection},
  author={Fang, Xiaocheng and Jin, Jiarui and Wang, Haoyu and Liu, Che and Cai, Jieyi and Nie, Guangkun and Li, Jun and Li, Hongyan and Hong, Shenda},
  journal={arXiv preprint arXiv:2509.19774},
  year={2025}
}

@inproceedings{pillaipapagei,
  title={PaPaGei: Open Foundation Models for Optical Physiological Signals},
  author={Pillai, Arvind and Spathis, Dimitris and Kawsar, Fahim and Malekzadeh, Mohammad},
  booktitle={The Thirteenth International Conference on Learning Representations}
}

@inproceedings{radford2021learning,
  title={Learning transferable visual models from natural language supervision},
  author={Radford, Alec and Kim, Jong Wook and Hallacy, Chris and Ramesh, Aditya and Goh, Gabriel and Agarwal, Sandhini and Sastry, Girish and Askell, Amanda and Mishkin, Pamela and Clark, Jack and others},
  booktitle={International conference on machine learning},
  pages={8748--8763},
  year={2021},
  organization={PmLR}
}

@article{kansal2025mc,
  title={MC-MED, multimodal clinical monitoring in the emergency department},
  author={Kansal, Aman and Chen, Emma and Jin, Boyang Tom and Rajpurkar, Pranav and Kim, David A},
  journal={Scientific Data},
  volume={12},
  number={1},
  pages={1094},
  year={2025},
  publisher={Nature Publishing Group UK London}
}

@article{johnson2016mimic,
  title={MIMIC-III, a freely accessible critical care database},
  author={Johnson, Alistair EW and Pollard, Tom J and Shen, Lu and Lehman, Li-wei H and Feng, Mengling and Ghassemi, Mohammad and Moody, Benjamin and Szolovits, Peter and Anthony Celi, Leo and Mark, Roger G},
  journal={Scientific data},
  volume={3},
  number={1},
  pages={1--9},
  year={2016},
  publisher={Nature Publishing Group}
}

@article{lee2022vitaldb,
  title={VitalDB, a high-fidelity multi-parameter vital signs database in surgical patients},
  author={Lee, Hyung-Chul and Park, Yoonsang and Yoon, Soo Bin and Yang, Seong Mi and Park, Dongnyeok and Jung, Chul-Woo},
  journal={Scientific Data},
  volume={9},
  number={1},
  pages={279},
  year={2022},
  publisher={Nature Publishing Group UK London}
}

@inproceedings{hong2020holmes,
  title={HOLMES: Health OnLine Model Ensemble Serving for Deep Learning Models in Intensive Care Units},
  author={Hong, Shenda and Xu, Yanbo and Khare, Alind and Priambada, Satria and Maher, Kevin and Aljiffry, Alaa and Sun, Jimeng and Tumanov, Alexey},
  booktitle={Proceedings of the 26th ACM SIGKDD International Conference on Knowledge Discovery \& Data Mining},
  pages={1614--1624},
  year={2020}
}

@article{pankaj2022review,
  title={A review on computation methods used in photoplethysmography signal analysis for heart rate estimation},
  author={Pankaj and Kumar, Ashish and Komaragiri, Rama and Kumar, Manjeet},
  journal={Archives of Computational Methods in Engineering},
  volume={29},
  number={2},
  pages={921--940},
  year={2022},
  publisher={Springer}
}

@article{wang2023pulsedb,
  title={PulseDB: A large, cleaned dataset based on MIMIC-III and VitalDB for benchmarking cuff-less blood pressure estimation methods},
  author={Wang, Weinan and Mohseni, Pedram and Kilgore, Kevin L and Najafizadeh, Laleh},
  journal={Frontiers in Digital Health},
  volume={4},
  pages={1090854},
  year={2023},
  publisher={Frontiers Media SA}
}

@article{chen2015racial,
  title={Racial/ethnic differences in sleep disturbances: the Multi-Ethnic Study of Atherosclerosis (MESA)},
  author={Chen, Xiaoli and Wang, Rui and Zee, Phyllis and Lutsey, Pamela L and Javaheri, Sogol and Alc{\'a}ntara, Carmela and Jackson, Chandra L and Williams, Michelle A and Redline, Susan},
  journal={Sleep},
  volume={38},
  number={6},
  pages={877--888},
  year={2015},
  publisher={Oxford University Press}
}

@misc{sun2023human,
  author       = {Sun, Haoqi and Ganglberger, Wolfgang and Nasiri, Samaneh and Gupta, Aditya and Ghanta, Manohar and Moura Junior, Valdery and Cash, Sydney and Stone, Katie and Zhang, Zhiyong and Ganjoo, Gauri and Nassi, Thijs E and Wei, Ruoqi and Meulenbrugge, Erik-Jan and Au, Rhoda and Clifford, Gari and Trotti, Lynn Marie and Hwang, Dennis and Mignot, Emmanuel and Katwa, Umakanth and Westover, M. Brandon},
  title        = {The Human Sleep Project (version 2.0)},
  year         = {2023},
  publisher    = {Brain Data Science Platform},
  doi          = {10.60508/qjbv-hg78},
  url          = {https://doi.org/10.60508/qjbv-hg78}
}

@article{redline1995familial,
  title={The familial aggregation of obstructive sleep apnea.},
  author={Redline, Susan and Tishler, Peter V and Tosteson, Tor D and Williamson, John and Kump, Kenneth and Browner, Ilene and Ferrette, Veronica and Krejci, Patrick},
  journal={American journal of respiratory and critical care medicine},
  volume={151},
  number={3},
  pages={682--687},
  year={1995},
  publisher={American Public Health Association}
}

@misc{reiss2019ppgdalia,
  author       = {Reiss, Attila and Indlekofer, Ina and Schmidt, Philip},
  title        = {PPG-DaLiA},
  year         = {2019},
  howpublished = {UCI Machine Learning Repository},
  doi          = {10.24432/C53890},
  url          = {https://doi.org/10.24432/C53890}
}

@misc{kachuee2015cuffless,
  author       = {Kachuee, Mohamad and Kiani, Mohammad and Mohammadzade, Hoda and Shabany, Mahdi},
  title        = {Cuff-Less Blood Pressure Estimation Dataset},
  year         = {2015},
  howpublished = {UCI Machine Learning Repository},
  doi          = {10.24432/C5B602},
  url          = {https://doi.org/10.24432/C5B602}
}

@article{nemcova2021brno,
  title={Brno university of technology smartphone ppg database (but ppg): Annotated dataset for ppg quality assessment and heart rate estimation},
  author={Nemcova, Andrea and Vargova, Enik{\"o} and Smisek, Radovan and Marsanova, Lucie and Smital, Lukas and Vitek, Martin},
  journal={BioMed Research International},
  volume={2021},
  number={1},
  pages={3453007},
  year={2021},
  publisher={Wiley Online Library}
}

@article{lee2018motion,
  title={Motion artifact cancellation in wearable photoplethysmography using gyroscope},
  author={Lee, H and Chung, H and Lee, J},
  journal={IEEE Sensors Journal},
  volume={19},
  number={3},
  pages={1166--1175},
  year={2018},
  publisher={IEEE}
}

@inproceedings{schmidt2018introducing,
  title={Introducing wesad, a multimodal dataset for wearable stress and affect detection},
  author={Schmidt, Philip and Reiss, Attila and Duerichen, Robert and Marberger, Claus and Van Laerhoven, Kristof},
  booktitle={Proceedings of the 20th ACM international conference on multimodal interaction},
  pages={400--408},
  year={2018}
}

@article{torres2020multi,
  title={Multi-task deep learning for cardiac rhythm detection in wearable devices},
  author={Torres-Soto, Jessica and Ashley, Euan A},
  journal={NPJ digital medicine},
  volume={3},
  number={1},
  pages={116},
  year={2020},
  publisher={Nature Publishing Group UK London}
}

@article{elgendi2012analysis,
  title={On the analysis of fingertip photoplethysmogram signals},
  author={Elgendi, Mohamed},
  journal={Current cardiology reviews},
  volume={8},
  number={1},
  pages={14--25},
  year={2012},
  publisher={Bentham science publishers}
}

@article{zhao2018sqi,
  title={SQI quality evaluation mechanism of single-lead ECG signal based on simple heuristic fusion and fuzzy comprehensive evaluation},
  author={Zhao, Zhidong and Zhang, Yefei},
  journal={Frontiers in physiology},
  volume={9},
  pages={727},
  year={2018},
  publisher={Frontiers Media SA}
}

@article{loshchilov2017fixing,
  title={Fixing weight decay regularization in adam},
  author={Loshchilov, Ilya and Hutter, Frank and others},
  journal={arXiv preprint arXiv:1711.05101},
  volume={5},
  number={5},
  pages={5},
  year={2017}
}

@article{yu2022coca,
  title={Coca: Contrastive captioners are image-text foundation models},
  author={Yu, Jiahui and Wang, Zirui and Vasudevan, Vijay and Yeung, Legg and Seyedhosseini, Mojtaba and Wu, Yonghui},
  journal={arXiv preprint arXiv:2205.01917},
  year={2022}
}

@article{orphanidou2017signal,
  title={Signal quality assessment in physiological monitoring: state of the art and practical considerations},
  author={Orphanidou, Christina},
  year={2017},
  publisher={Springer}
}

@inproceedings{chen2020simple,
  title={A simple framework for contrastive learning of visual representations},
  author={Chen, Ting and Kornblith, Simon and Norouzi, Mohammad and Hinton, Geoffrey},
  booktitle={International conference on machine learning},
  pages={1597--1607},
  year={2020},
  organization={PmLR}
}

@article{grill2020bootstrap,
  title={Bootstrap your own latent-a new approach to self-supervised learning},
  author={Grill, Jean-Bastien and Strub, Florian and Altch{\'e}, Florent and Tallec, Corentin and Richemond, Pierre and Buchatskaya, Elena and Doersch, Carl and Avila Pires, Bernardo and Guo, Zhaohan and Gheshlaghi Azar, Mohammad and others},
  journal={Advances in neural information processing systems},
  volume={33},
  pages={21271--21284},
  year={2020}
}

@article{ansari2025chronos,
  title={Chronos-2: From univariate to universal forecasting},
  author={Ansari, Abdul Fatir and Shchur, Oleksandr and K{\"u}ken, Jaris and Auer, Andreas and Han, Boran and Mercado, Pedro and Rangapuram, Syama Sundar and Shen, Huibin and Stella, Lorenzo and Zhang, Xiyuan and others},
  journal={arXiv preprint arXiv:2510.15821},
  year={2025}
}

@article{goswami2024moment,
  title={Moment: A family of open time-series foundation models},
  author={Goswami, Mononito and Szafer, Konrad and Choudhry, Arjun and Cai, Yifu and Li, Shuo and Dubrawski, Artur},
  journal={arXiv preprint arXiv:2402.03885},
  year={2024}
}

@misc{siam2019real,
  title = {Real-World {PPG} dataset},
  author = {Siam, Ali and Abd El-Samie, Fathi and Abu Elazm, Atef and El-Bahnasawy, Nirmeen and Elbanby, Ghada},
  year = {2019},
  publisher = {Mendeley Data},
  version = {V1},
  doi = {10.17632/yynb8t9x3d.1},
  url = {https://data.mendeley.com/datasets/yynb8t9x3d/1}
}

@article{makowski2021neurokit2,
  title={NeuroKit2: A Python toolbox for neurophysiological signal processing},
  author={Makowski, Dominique and Pham, Tam and Lau, Zen J and Brammer, Jan C and Lespinasse, Fran{\c{c}}ois and Pham, Hung and Sch{\"o}lzel, Christopher and Chen, SH Annabel},
  journal={Behavior research methods},
  volume={53},
  number={4},
  pages={1689--1696},
  year={2021},
  publisher={Springer}
}

@article{wang2025reliability,
  title={Reliability and validity of a novel single-lead portable electrocardiogram device for pregnant women: a comparative study},
  author={Wang, Haixue and Wang, Jianwei and Jing, Wei and Dai, Shanshan and Zhang, Deyun and Geng, Shijia and Wang, Haijun and Hong, Shenda},
  journal={BMC Medical Informatics and Decision Making},
  volume={25},
  number={1},
  pages={108},
  year={2025},
  publisher={Springer}
}

@article{valenza2025brain,
  title={The brain--heart axis: Integrative cooperation of neural, mechanical and biochemical pathways},
  author={Valenza, Gaetano and Mati{\'c}, Zoran and Catrambone, Vincenzo},
  journal={Nature Reviews Cardiology},
  pages={1--14},
  year={2025},
  publisher={Nature Publishing Group UK London}
}

@article{duh2017diabetic,
  title={Diabetic retinopathy: current understanding, mechanisms, and treatment strategies},
  author={Duh, Elia J and Sun, Jennifer K and Stitt, Alan W},
  journal={JCI insight},
  volume={2},
  number={14},
  pages={e93751},
  year={2017}
}

@inproceedings{erturk2025beyond,
  title={Beyond Sensor Data: Foundation Models of Behavioral Data from Wearables Improve Health Predictions},
  author={Erturk, Eray and Kamran, Fahad and Abbaspourazad, Salar and Jewell, Sean and Sharma, Harsh and Li, Yujie and Williamson, Sinead and Foti, Nicholas J and Futoma, Joseph},
  booktitle={International Conference on Machine Learning},
  pages={15516--15541},
  year={2025},
  organization={PMLR}
}

@inproceedings{narayanswamyscaling,
  title={Scaling Wearable Foundation Models},
  author={Narayanswamy, Girish and Liu, Xin and Ayush, Kumar and Yang, Yuzhe and Xu, Xuhai and Garrison, Jake and Tailor, Shyam A and Sunshine, Jacob and Liu, Yun and Althoff, Tim and others},
  booktitle={The Thirteenth International Conference on Learning Representations}
}

@article{fang2026ecgflowcmr,
  title={ECGFlowCMR: Pretraining with ECG-Generated Cine CMR Improves Cardiac Disease Classification and Phenotype Prediction},
  author={Fang, Xiaocheng and Ding, Zhengyao and Cai, Jieyi and Xiao, Yujie and Liu, Bo and Jin, Jiarui and Wang, Haoyu and Nie, Guangkun and Huang, Shun and Chen, Ting and others},
  journal={arXiv preprint arXiv:2601.20904},
  year={2026}
}

@article{zhang2026ecgomics,
  title={ECGomics: An open platform for AI-ECG digital biomarker discovery},
  author={Zhang, Deyun and Li, Jun and Geng, Shijia and Wang, Yue and Chen, Shijie and Fan, Sumei and Zhao, Qinghao and Hong, Shenda},
  journal={Health Data Science},
  volume={6},
  pages={0427},
  year={2026},
  publisher={AAAS}
}

@article{yin2024multi,
  title={Multi-modal clip-informed protein editing},
  author={Yin, Mingze and Zhou, Hanjing and Zhu, Yiheng and Lin, Miao and Wu, Yixuan and Wu, Jialu and Xu, Hongxia and Hsieh, Chang-Yu and Hou, Tingjun and Chen, Jintai and others},
  journal={Health Data Science},
  volume={4},
  pages={0211},
  year={2024},
  publisher={AAAS}
}

@article{he2024exploring,
  title={Exploring unlabeled data in multiple aspects for semi-supervised MRI segmentation},
  author={He, Qingyuan and Yan, Kun and Luo, Qipeng and Yi, Duan and Wang, Ping and Han, Hongbin and Liu, Defeng},
  journal={Health Data Science},
  volume={4},
  pages={0166},
  year={2024},
  publisher={AAAS}
}

\end{document}